\documentclass[showpacs,preprintnumbers,amsmath,amssymb]{revtex4}
\usepackage{graphicx}
\usepackage{subfigure}
\begin{document}

\title{Radiative Corrections to Higgs Masses in $Z^{\prime}$ Models}
\author{Hassib Amini}
\affiliation{School of Physics and Astronomy, University of Minnesota, Minneapolis, MN 55455}

\begin{abstract}
We calculate radiative corrections to the masses of the Higgs bosons
in a minimal supersymmetric model that contains an additional non-anomalous
\( U(1)' \) gauge symmetry. With some fine-tuning of the \( U(1)' \)
charges of the Higgs fields, it is possible to suppress the \( Z \)--\( Z^{\prime } \)
mixing. We use this fact, along with the lower bound on the lightest
Higgs mass after LEPII era, as a criterion to restrict the set of
parameters in our analysis. We calculate the mass of the lightest
Higgs and its mixing with the other Higgs bosons, in a large region
of the parameter space. 
\end{abstract}

\maketitle

\section{Introduction}

Supersymmetry (SUSY) is one of the most viable extensions of the Standard
Model (SM) for curing the quantum instability of the Higgs sector.
SUSY, however, has a new hierarchy problem concerning the natural
scale of the Higgsino Dirac mass -- the \( \mu  \) problem. Many
supersymmetric extensions of SM have been proposed to cure the \( \mu  \)
problem. Although there is no natural solution to this problem in
the minimal supersymmetric models, the next-to-minimal SUSY \cite{ellis1}
solves the \( \mu  \) problem, but at the expense of inducing large-tension
domain walls, which can over-close the universe. Among all possible
extensions of supersymmetric SM, however, the ones that invoke an
extra \( U(1)' \) symmetry, which forbids a bare \( \mu  \) term
\cite{kim}, are particularly promising since problems with the generation
of tadpoles and axion are automatically avoided. 
Also, extra \( U(1)' \) gauge symmetries, with their associated extra \( Z^\prime \) bosons,
naturally arise in effective theories coming from breaking of 
grand unified theories (GUTs) or in superstring
compactifications \cite{dine}. The above two reasons are the major
motivations for our paper.

In general, the presence of a \( Z' \) boson in the low-energy spectrum
generates additional neutral current transitions, which can have important
implications for precision tests \cite{amaldi} or flavor violation
\cite{langacker}. It has been shown that the existing data allow
for a light \( Z' \) boson with family-dependent couplings \cite{erler}.
In a general \( Z^{\prime } \) model, the hierarchy of the soft masses
in the Higgs sector leads to different vacua in which the \( Z^{\prime } \)
boson can be heavy or light, although the \( Z \)--\( Z^{\prime } \)
mixing is sufficiently small in all cases. The analysis of Erler and
Langacker \cite{erler} shows that the \( Z^{\prime } \) boson can
be as light as \( 200\, \, {\textrm{GeV}} \) with a \( Z \)--\( Z^{\prime } \)
mixing angle of the \( {\mathcal{O}}(10^{-3}) \).

Although a low-scale extra \( U(1)' \) symmetry stabilizes the \( \mu  \)
parameter to TeV scale, the tree-level Higgs potential is far from
representing the physical observables with sufficient precision. In
fact, experience from the minimal model shows that there are large
corrections to the Higgs masses, once the radiative effects are taken
into account \cite{ellis2}. This is also the case with the \( Z' \)
models, as the partial analysis of the Higgs sector shows \cite{daikoku}.
Therefore, for a proper analysis of the collider data concerning the
Higgs production, it is necessary to compute the Higgs masses and
couplings at least at the one-loop level. In fact, this is necessary
even for making comparisons or putting constraints on the parameter
space using the LEP results \cite{lep}.

Our main goal in this paper is to work out the neutral Higgs boson
masses and mixing angles at the one-loop level in supergravity
models with an additional \( U(1)' \) gauge symmetry. A general tree-level
analysis of such models can be found in \cite{cvetic}. We will assume
that the parameters of the model have already been fixed to the TeV
scale via the one-loop renormalization group equation (RGE) 
running from the string scale with appropriate
initial conditions. The low energy mass spectrum and appropriate vacua
that naturally suppress the \( Z \)--\( Z^{\prime } \) mixing angle
have already been derived in \cite{cvetic} for the tree-level Higgs potential.
We will further assume that there is no violation of the CP invariance
in the Higgs sector, which is a simplifying assumption rather than
a result following from high--scale model building.

In the next section we review the properties of the tree-level potential.
In Sec. 3, we compute the effective potential and derive the Higgs
mass spectrum at the one-loop level. We will also briefly comment on the
relevance of two-loop effects.
In Sec. 4, we present some numerical
estimates of the masses and mixing angles for likely values of the
parameter values. Here, the lower bound set on the lightest Higgs
mass by LEPII is used as a constraint to restrict the values of our
parameters. Of course, to improve the bounds on our parameter space
and to further restrict the parameter values, we need to consider
the Higgs production and decay rates. The dominant channels are the
Higgs-strahlung (Bjorken) process, 
\( e^{-}e^{+}\rightarrow Z_{i}^{*}\rightarrow Z_{j}h_{k} \),
and the \( Z \) decay into two Higgses, 
\( e^{-}e^{+}\rightarrow Z_{i}^{*}\rightarrow h_{j}h_{k} \).
However, we will work out the details of these processes and cross-sections
in a separate work. For now, the bounds that we obtain using the lightest
Higgs mass as a constraint are still valid. Finally, in Sec.5, we
conclude the work and discuss its implications for Higgs phenomenology.

\section{Tree-Level Effective Potential}

We first describe the structure of the effective potential at the
tree level. The gauge group is the same as that of the Standard Model,
but with an additional \( U(1)' \) factor, i.e., 
\( G=SU(3)_{c}\times SU(2)_{L}\times U(1)_{Y}\times U(1)_{Y'} \),
with coupling constants \( g_{3} \), \( g_{2} \), \( g_{1} \),
and \( g_{1}' \), respectively. The Higgs sector contains two Higgs
doublets \( H_{1} \) and \( H_{2} \), and one singlet \( S \).
The matter multiplets are given by left-handed chiral superfields.
For specific charge assignments of the chiral superfields with respect
to the gauge group, we refer the readers to Ref. \cite{cvetic}. For
our purposes, we only need to know the charges of the Higgs fields
under the extra \( U(1)' \) gauge symmetry. We denote the Higgs charges
by \( Q_{1} \), \( Q_{2} \), and \( Q_{s} \). The superpotential
is given by 
\begin{eqnarray}
\mathcal{W} & = & h_{s}\hat{S}(\hat{H}_{1}\cdot \hat{H}_{2})+h_{t}\hat{U}^{c}(\hat{Q}\cdot \hat{H}_{2}).\label{Eq1} 
\end{eqnarray}
 The most important feature of the above superpotential that distinguishes
it from either MSSM or NMSSM, is the absence of a cubic term in \( \hat{S} \)
and a term proportional to \( \hat{H}_{1}\cdot \hat{H}_{2} \), usually called
the \( \mu  \) term %
\cite{footnote}. The gauge invariance of the superpotential under the \( U(1)' \) forbids
the appearance of such terms. Although a \( \mu  \) term is absent
from the superpotential, an effective \( \mu  \) parameter is generated
by the vacuum expectation value of the scalar field \( S \). We use the hatted fields to
denote chiral Higgs superfields, and the un-hatted fields to denote scalar Higgs fields.

We parameterize the explicit soft breaking of supersymmetry by 
\begin{eqnarray}
{\mathcal{L}}_{SB} & = & -\sum _{i}M_{i}\tilde{\lambda} _{i}\tilde{\lambda} _{i}+Ah_{s}SH_{1}\cdot
H_{2}+A_{t}h_{t}\tilde{U}^{c}\tilde{Q}\cdot H_{2}\nonumber \\
 & - & m_{Q}^{2}\left| \tilde{Q}\right| ^{2}-m_{U}^{2}\left| \tilde{U}\right| ^{2}-m_{D}^{2}\left| \tilde{D}\right|
 ^{2}-m_{E}^{2}\left| \tilde{E}\right| ^{2}\nonumber \\
 & - & m_{L}^{2}\left| \tilde{L}\right| ^{2}-m_{1}^{2}\left| H_{1}\right| ^{2}-m_{2}^{2}\left| H_{2}\right| ^{2}-m_{s}^{2}\left| S\right| ^{2},\label{Eq2} 
\end{eqnarray}
 where \( \tilde{\lambda} _{i} \) are the gaugino fields, and the hermitian conjugate terms are assumed to keep the potential real. 
The tree-level Higgs potential follows from \( {\mathcal{L}}_{SB} \), \( F \) and
\( D \) terms:
\begin{widetext}
\begin{eqnarray}
V_{F} & = & \left| h_{s}\right| ^{2}\left[ \left| H_{1}\cdot H_{2}\right| ^{2}+\left| S\right| ^{2}\left( \left| H_{1}\right| ^{2}+\left| H_{2}\right| ^{2}\right) \right] ,\nonumber \\
V_{D} & = & \frac{g_{1}^{2}+g_{2}^{2}}{8}\left[ \left| H_{2}\right| ^{2}-\left| H_{1}\right| ^{2}\right] ^{2}+\frac{g_{2}^{2}}{2}\left| H_{1}^{\dagger }H_{2}\right| ^{2}+\frac{g_{1}^{'2}}{2}\left[ Q_{1}\left| H_{1}\right| ^{2}+Q_{2}\left| H_{2}\right| ^{2}+Q_{s}\left| S\right| ^{2}\right] ^{2},\nonumber \\
V_{S} & = & m_{1}^{2}\left| H_{1}\right| ^{2}+m_{2}^{2}\left| H_{2}\right| ^{2}+m_{s}^{2}\left| S\right| ^{2}-\left[ Ah_{s}SH_{1}\cdot H_{2}+h.c.\right] ,\label{Eq3} 
\end{eqnarray}
\end{widetext}
 where \( Q_{i} \) are the \( U(1)' \) charges of \( H_{i} \) and
\( Q_{s}+Q_{1}+Q_{2}=0 \) due to the gauge invariance of the superpotential
under the extra gauge symmetry. Above we notice that \( g_{1}' \)
always appears in combination with \( Q_{i} \). Hence, it is convenient
to absorb it in the definition of \( Q_{i} \) and define new charges
\( Q_{i}=g_{1}'Q_{i} \). Therefore, \( g_{1}' \) will not explicitly
appear in our formulae unless stated otherwise.

We assume that all the coupling constants in the above potential are
real. At the tree level, the potential cannot violate CP symmetry
either explicitly or spontaneously. A possible phase could come from
\( Ah_{s} \); however, such a phase could be absorbed into the global
phases of the Higgs fields. At the one-loop level, CP symmetry can
be explicitly broken due to the complex phases in the scalar quark sector.
However, we set all the CP-violating phases equal to zero and consider
only the CP-conserving scenario.

The Higgs sector of the theory contains ten real degrees of freedom.
Each Higgs doublet contains four real fields and the Higgs singlet
contains two real fields. After the electroweak symmetry breaking, four
of the ten fields become the longitudinal components of the four vector
bosons in our model. The remaining six fields result in three scalars,
one pseudoscalar, and one charged Higgs. We decompose the Higgs fields
as 
\begin{eqnarray*}
H_{1}=\left[ \begin{array}{c}
\phi _{1}^{0}\\
\phi _{1}^{-}
\end{array}\right] ,\qquad H_{2}=\left[ \begin{array}{c}
\phi _{2}^{+}\\
\phi _{2}^{0}
\end{array}\right] , &  & 
\end{eqnarray*}
 where the neutral components will be further separated into scalar
and pseudoscalar bosons below.

The vacuum state of the theory is defined by the Higgs vacuum expectation
values (VEVs): 
\( \left\langle \phi _{1}^{-}\right\rangle =\left\langle \phi _{2}^{+}\right\rangle =0 \),
\( \left\langle \phi _{i}^{0}\right\rangle =v_{i}/\sqrt{2} \), and
\( \left\langle S\right\rangle =v_{3}/\sqrt{2} \), where \( v_{i} \) are real, 
and \( v^{2}=v_{1}^{2}+v_{2}^{2}=(246\, \, {\textrm{GeV}})^{2} \).
The effective \( \mu \) parameter is generated by the VEV of \( S \), and is defined by
\( \mu_{s} = h_{s}v_{3}/\sqrt{2} \). 
Here \( v_{1}=v\cos \beta  \), \( v_{2}=v\sin \beta  \), and \( v_{3} \equiv x \). 
For this to be a physical minimum, the potential
must be negative when evaluated at the point \( (v_{1},v_{2},x) \),
and the masses of the Higgs bosons must be positive. Even when these
conditions are satisfied, the above point is not guaranteed to be
the absolute minimum. Whether it is still acceptable depends on the
location and depth of the other minima and the width between them.
At the minimum point, the potential has vanishing first derivatives
with respect to the three CP-even scalars, \( i.e. \), all tadpoles
vanish. This enables one to trade the soft mass-squared parameters for their VEVs.

The tree-level masses of the Higgs bosons are obtained by diagonalizing
their corresponding field-dependent mass-squared matrices. To this
end, we need to substitute 
\( \phi _{i}^{0}=\left( v_{i}+\phi _{i}+i\varphi _{i}\right) /\sqrt{2} \)
and \( S=\left( x+\phi _{3}+i\varphi _{3}\right) /\sqrt{2} \) into
the potential. Here \( \phi _{i} \) and \( \varphi _{i} \) stand
for CP-even and CP-odd directions, respectively. Using the basic
definitions 
\begin{eqnarray}
\left( {\mathcal{M}}_{S}^{0}\right) _{ij}=\left\langle \frac{\partial ^{2}V^{0}}{\partial \phi _{i}\partial \phi
_{j}}\right\rangle ,\qquad \left( {\mathcal{M}}_{P}^{0}\right) _{ij}=\left\langle \frac{\partial ^{2}V^{0}}{\partial \varphi
_{i}\partial \varphi _{j}}\right\rangle , &  & \label{Eq4} 
\end{eqnarray}
 we form the Higgs mass-squared matrix. Evaluation of Eq. (\ref{Eq4})
at the tree level is straight forward. Defining \( m_{3}^2 = A\mu_{s} \), we obtain
\begin{eqnarray}
\left( {\mathcal{M}}_{S}^{0}\right) _{11} & = & \left[ Q_{1}^{2}+\frac{g_{1}^{2}+g_{2}^{2}}{4}\right] v_{1}^{2}+m_{3}^{2}\tan \beta ,\nonumber \\
\left( {\mathcal{M}}_{S}^{0}\right) _{12} & = & \left[ h_{s}^{2}+Q_{1}Q_{2}-\frac{g_{1}^{2}+g_{2}^{2}}{4}\right] v_{1}v_{2}-m_{3}^{2},\nonumber \\
\left( {\mathcal{M}}_{S}^{0}\right) _{13} & = & \left[ h_{s}^{2}+Q_{1}Q_{s}\right] v_{1}x-m_{3}^{2}\frac{v\sin \beta }{x},\nonumber \\
\left( {\mathcal{M}}_{S}^{0}\right) _{22} & = & \left[ Q_{2}^{2}+\frac{g_{1}^{2}+g_{2}^{2}}{4}\right] v_{2}^{2}+m_{3}^{2}\cot \beta ,\nonumber \\
\left( {\mathcal{M}}_{S}^{0}\right) _{23} & = & \left[ h_{s}^{2}+Q_{2}Q_{s}\right] v_{2}x-m_{3}^{2}\frac{v\cos \beta }{x},\nonumber \\
\left( {\mathcal{M}}_{S}^{0}\right) _{33} & = & \left[ Q_{s}^{2}\right] x^{2}+m_{3}^{2}\frac{v^{2}\sin \beta \cos \beta
}{x^{2}}.\label{Eq5} 
\end{eqnarray}
 For the pseudoscalar mass-squared matrix, we get 
\begin{equation}
\left( \mathcal{M}_{P}^{0} \right)_{ij} = \frac{h_{s}A}{\sqrt{2}} \frac{v_{1}v_{2}v_{3}}{v_{i}v_{j}}. \label{Eq6} 
\end{equation}
 The eigenvalues of the scalar mass-squared matrix correspond to the
masses of the Higgs bosons. Although these eigenvalues can be obtained
analytically, they are often too complicated to be useful. However,
from the structure of the above matrix we can obtain useful information
about its smallest eigenvalue, which corresponds to the lightest Higgs
mass. Namely, for any symmetric \( n\times n \) matrix, its smallest
eigenvalue is less than the smaller eigenvalue of it left upper \( 2\times 2 \)
sub-matrix. With this observation, we get 
\begin{equation}
\left( M_{h_{1}}^{0}\right) ^{2}\leq M_{Z}^{2}\cos ^{2}2\beta +\frac{1}{2}h_{s}^{2}v^{2}\sin ^{2}2\beta +Q_{H}^{2}v^{2},\label{Eq7}
\end{equation}
where \( Q_{H} = Q_{1} \cos^{2} \beta + Q_{2} \sin^{2} \beta \). 
The first two terms are familiar from NMSSM \cite{ellis1}, while
the third term is unique to the model under consideration. Notice
that this term allows the lightest Higgs mass to be larger than that
predicted by either MSSM or NMSSM.

After appropriate rotations, the pseudoscalar mass-squared matrix
gives one non-zero eigenvalue corresponding to the physical pseudoscalar
mass. The other two eigenvalues which are zero, are the Goldstone
degrees of freedom. After electroweak symmetry breaking, these become
the longitudinal components of \( Z \) and \( Z' \). The \( Z \)--\( Z^{\prime } \)
mass-squared matrix is given by 
\begin{eqnarray}
\left( {\mathcal{M}}\right) _{Z-Z'} & = & \left[ \begin{array}{ll}
M_{Z}^{2} & \triangle ^{2}\\
\triangle ^{2} & M_{Z'}^{2}
\end{array}\right] ,\label{Eq8} 
\end{eqnarray}
 where \( M_{Z'}^{2}=(Q_{1}^{2}v_{1}^{2}+Q_{2}^{2}v_{2}^{2}+Q_{s}^{2}x^{2}) \)
and \( \triangle ^{2}=(1/2)\sqrt{g_{1}^{2}+g_{2}^{2}}\left( v_{1}^{2}Q_{1}-v_{2}^{2}Q_{2}\right)  \).
The eigenvalues of the above matrix, together with the \( Z \)--\( Z^{\prime } \)
mixing angle, are given by 
\begin{eqnarray}
M_{Z_{1,2}} ^{2} & = & \frac{1}{2}\left[ M_{Z}^{2}+M_{Z'}^{2}\mp \sqrt{\left( M_{Z}^{2}-M_{Z'}^{2}\right) ^{2}+4\triangle ^{4}}\right] ,\nonumber \\
\alpha _{Z-Z'} & = & \frac{1}{2}\arctan \left[ \frac{2\triangle ^{2}}{M_{Z'}^{2}-M_{Z}^{2}}\right] .\label{Eq9} 
\end{eqnarray}
 The mixing angle \( \alpha _{Z-Z'} \) has to be smaller than a few
times \( 10^{-3} \), so that \( M_{Z_{1}} \) would correspond to
the observed \( Z \) boson mass. For completeness, we also give the
expressions for the pseudoscalar mass and the charged Higgs mass:
\begin{eqnarray}
\left( M_{P}^{0}\right) ^{2} & = & \frac{A\mu _{s}}{\sin \beta \cos \beta }\left[ 1+\frac{v^{2}\sin ^{2}\beta \cos ^{2}\beta }{x^{2}}\right] ,\nonumber \\
\left( M_{H^{\pm }}\right) ^{2} & = & M_{W}^{2}+\frac{A\mu _{s}}{\sin \beta \cos \beta }-\frac{1}{2}h_{s}v^{2}.\label{Eq10} 
\end{eqnarray}
 From the above, it is clear that the pseudoscalar mass is never negative,
while the charged Higgs mass can be lower than the \( W \) boson
mass, and can even run to negative values for some choices of the parameters.

In the next section, we include the main one-loop contributions to
the tree-level effective potential. In general, the tree-level potential
is written in terms of the running coupling constants and masses, which
are defined at some renormalization point \( Q \). However, the tree-level 
effective potential written in terms of running parameters is
too sensitive to the choice of \( Q \), and one cannot make reliable
calculations. The situation is considerably improved when one includes
the one-loop contributions to the effective potential \cite{Appelquist}.
We take into account the one-loop top/stop and sbottom effects, which 
are the main corrections.

\section{One-Loop Effective Potential}

As explained at the end of the previous section, the most important
one-loop contribution to the tree-level effective potential comes
from the top and scalar top quarks. However, the contribution of the
bottom scalar quarks can also be sizable when \( \tan \beta \sim 40 \)
or larger. We take both contributions into account. For the rest of this section,
we will state our results in full generality, making no assumptions
about the numerical values of our parameters.

The stop and sbottom mass-squared matrices are given by 
\begin{eqnarray}
\left( \mathcal{M}_{t} \right) _{11} & = & m_{Q}^{2}+k_{1}^{t}\left| H_{1}^{0}\right|
^{2}+k_{2}^{t}\left|H_{2}^{0}\right|^{2}+k_{s}^{t}\left| S\right| ^{2} , \nonumber \\ 
\left( \mathcal{M}_{t} \right) _{12} & = & h_{t}(A_{t}H_{2}^{0*}-h_{s}SH_{1}^{0}) , \nonumber \\
\left( \mathcal{M}_{t} \right) _{22} & = & m_{U}^{2}+l_{1}^{t}\left| H_{1}^{0}\right| ^{2}+l_{2}^{t}\left|H_{2}^{0}\right|^{2}+l_{s}^{t}\left| S\right| ^{2} , \nonumber \\
\left( \mathcal{M}_{b} \right) _{11} & = & m_{Q}^{2}+k_{1}^{b}\left| H_{1}^{0}\right| ^{2}+k_{2}^{b}\left|
H_{2}^{0}\right|^{2}+k_{s}^{b}\left| S\right| ^{2} , \nonumber \\ 
\left( \mathcal{M}_{b} \right) _{12} & = & h_{b}(h_{s}SH_{2}^{0}-A_{b}H_{1}^{0*}) , \nonumber \\
\left( \mathcal{M}_{b} \right) _{22} & = & m_{D}^{2}+l_{1}^{b}\left| H_{1}^{0}\right|
^{2}+l_{2}^{b}\left|H_{2}^{0}\right|^{2}+l_{s}^{b}\left| S\right| ^{2} , \label{Eq11}
\end{eqnarray}
 where \( k^{t} _{1}=\left( g_{2}^{2}-g_{1}^{2}/3\right) /4+Q_{Q}Q_{1} \),
\( k^{t} _{2}=h_{t}^{2}-\left( g_{2}^{2}-g_{1}^{2}/3\right) /4+Q_{Q}Q_{2} \),
\( k^{t} _{s}=Q_{Q}Q_{s} \), \( l^{t} _{1}=g_{1}^{2}/3+Q_{U}Q_{1} \),
\( l^{t} _{2}=h_{t}^{2}-g_{1}^{2}/3+Q_{U}Q_{1} \), \( l^{t} _{s}=Q_{U}Q_{s} \), and
\( k^{b} _{1}=h_{b}^{2}-\left(g_{2}^{2}+g_{1}^{2}/3\right) /4+Q_{Q}Q_{1} \),
\( k^{b} _{2}=\left( g_{2}^{2}+g_{1}^{2}/3\right) /4+Q_{Q}Q_{2} \),
\( k^{b} _{s}=Q_{Q}Q_{s} \), \( l^{b} _{1}=h_{b}^{2}-g_{1}^{2}/6+Q_{D}Q_{1} \),
\( l^{b} _{2}=g_{1}^{2}/6+Q_{D}Q_{1} \), and finally, \( l^{b} _{s}=Q_{D}Q_{s} \).
The cancellation of triangle anomalies gives \( Q_{Q}=-Q_{1}/3 \), 
\( Q_{U}=\left( Q_{1}-2Q_{2}\right) /3 \), and \( Q_{D}=\left( Q_{1}+2Q_{2}\right) /3 \). 
The eigenvalues of the above matrix are the masses of the left-handed and right-handed
stops and sbottoms, given by 
\begin{eqnarray}
m_{\tilde{q}_{1,2}}^{2} & = & \frac{1}{2}tr{\mathcal{M}}_{q}\mp
\frac{1}{2}\sqrt{(tr{\mathcal{M}}_{q})^{2}-4det{\mathcal{M}}_{q}}.\label{Eq12} 
\end{eqnarray}
 Using the stop and sbottom masses from above, we can express the one-loop correction
to the effective potential by the Coleman-Weinberg formula \cite{coleman}
\begin{eqnarray}
V^{1} & = &  k\left[ m_{\tilde{q}_{j}}^{4}\left( \log \frac{m_{\tilde{q}_{j}}^{2}}{Q^{2}}-\frac{3}{2}\right)
-2\bar{m}_{q}^{4}\left( \log \frac{\bar{m}_{q}^{2}}{Q^{2}}-\frac{3}{2}\right) \right], \nonumber \\
\label{Eq13} 
\end{eqnarray}
 where \( Q \) is the renormalization scale in the \( \overline{MS} \)
scheme and \( k=3/(32\pi ^{2}) \). We sum over \( q=(t,b) \) and \( j=(1,2) \).
Here, \( \bar{m}_{t}^{2} = h_{t}^{2}\left| H_{2}^{0}\right| ^{2}\) and 
\( \bar{m}_{b}^{2} = h_{b}^{2}\left| H_{1}^{0}\right| ^{2}\). The one-loop scalar and
pseudoscalar mass-squared matrices are given by 
\begin{eqnarray}
\left( {\mathcal{M}}_{S}^{1}\right) _{ij}=\left\langle \frac{\partial ^{2}V^{1}}{\partial \phi _{i}\partial \phi _{j}}-\delta
_{ij}\frac{1}{\phi _{i}}\frac{\partial V^{1}}{\partial \phi _{j}}\right\rangle ,\nonumber\\
\left( {\mathcal{M}}_{P}^{1}\right) _{ij}=\left\langle \frac{\partial ^{2}V^{1}}{\partial \varphi _{i}\partial \varphi
_{j}}-\delta _{ij}\frac{1}{\phi _{i}}\frac{\partial V^{1}}{\partial \phi _{j}}\right\rangle , &  & \label{Eq14} 
\end{eqnarray}
 where the second term in the brackets is due to the fact that the position
of the minimum has shifted because of the one-loop effects. By substituting
Eq. (\ref{Eq12}) into Eq. (\ref{Eq13}) and substituting the resulting
expression into Eq. (\ref{Eq14}), we get 
\begin{widetext}
\begin{eqnarray}
({\mathcal{M}}_{S}^{1})_{ij} & = & \frac{3}{32\pi ^{2}} \sum_{q=t,b}\left[ \xi _{ij} ^{q}{\mathcal{F}}\left(
m_{\tilde{q}_{1}}^{2},m_{\tilde{q}_{2}}^{2}\right) +\zeta _{ij} ^{q}{\mathcal{G}}\left(
m_{\tilde{q}_{1}}^{2},m_{\tilde{q}_{2}}^{2}\right) +\rho _{ij} ^{q}\ln \left(
\frac{m_{\tilde{q}_{1}}^{2}}{m_{\tilde{q}_{2}}^{2}}\right) -\lambda _{ij} ^{q}\ln \left( \frac{m_{q}^{4}}{Q^{4}}\right) \right] ,\nonumber \\
({\mathcal{M}}_{P}^{1})_{ij} & = & \frac{3}{32\pi ^{2}}\sum_{q=t,b}\left[ \eta _{ij}^{q}{\mathcal{F}}\left(
m_{\tilde{q}_{1}}^{2},m_{\tilde{q}_{2}}^{2}\right) \right],\label{Eq15} 
\end{eqnarray}
where
\begin{eqnarray}
\xi _{ij} ^{q} & = & v_{i}v_{j}\left( k _{i} ^{q}k _{j} ^{q}+l _{i} ^{q}l _{j} ^{q}+
\delta_{i1(2)} \delta_{j3} h_{s}^{2}h_{q}^{2}\right)+\left( -1\right) ^{1-\delta _{ij}}\frac{h_{s}A_{q}h_{q}^{2}}{\sqrt{2}}\frac{v_{1}v_{2}v_{3}}{v_{i}v_{j}},\nonumber \\
\zeta _{ij} ^{q} & = & \frac{1}{2}v_{i}v_{j}\left[ R_{i} ^{q}R_{j} ^{q}+\left( k _{i}^{q}+l _{i}^{q}\right) \left( k _{j}^{q}+l _{j}^{q}\right) \right] ,\nonumber \\
\rho _{ij} ^{q} & = & v_{i}v_{j}\left[ \left( k _{j}^{q}+l _{j}^{q}\right) R_{i}+\left( k _{i}^{q}+l _{i}^{q}\right) R_{j}^{q}\right] ,\nonumber \\
\eta _{ij} ^{q} & = & \frac{h_{s}A_{q}h_{q}^{2}}{\sqrt{2}}\frac{v_{1}v_{2}v_{3}}{v_{i}v_{j}}, \nonumber\\
R_{1}^{q} & = & \frac{\sqrt{2}h_{s}A_{q}h_{q}^{2}x\tan \beta
-h_{s}^{2}h_{t}^{2}x^{2}(2h_{b}^{2}A_{b}^{2})-m_{Q}^{2}-m_{U(D)}^{2}-\left( k_{1}^{q}-l_{1}^{q}\right) \sum_{j=1}^{3} \left( k_{j}^{q}-l_{j}^{q}\right) v_{j}}{m_{\tilde{q}_{2}}^{2}-m_{\tilde{q}_{1}}^{2}},\nonumber \\
R_{2}^{q} & = & \frac{\sqrt{2}h_{s}A_{q}h_{q}^{2}x\cot \beta
-2h_{t}^{2}A_{t}^{2}(h_{s}^{2}h_{b}^{2}x^{2})-m_{Q}^{2}-m_{U(D)}^{2}-\left( k_{2}^{q}-l_{2}^{q}\right) \sum_{j=1}^{3} \left( k_{j}^{q}-l_{j}^{q}\right) v_{j}}{m_{\tilde{q}_{2}}^{2}-m_{\tilde{q}_{1}}^{2}},\nonumber \\
R_{3}^{q} & = & \frac{\sqrt{2}h_{s}A_{q}h_{q}^{2}v_{1}v_{2}/x-h_{s}^{2}h_{q}^{2}v_{2}^{2}-m_{Q}^{2}-m_{U(D)}^{2}-\left(
k_{3}^{q}-l_{3}^{q}\right) \sum_{j=1}^{3} \left( k_{j}^{q}-l_{j}^{q}\right) v_{j}}{m_{\tilde{q}_{2}}^{2}-m_{\tilde{q}_{1}}^{2}},
\label{Eq17} 
\end{eqnarray}
\end{widetext}
where \( q=t \) \( (b) \) for top (bottom) quark/squarks.
Above, \( \lambda _{22} ^{t} = 2h_{t}^{4}v_{2}v_{2} \), \( \lambda _{11} ^{b} = 2h_{b}^{4}v_{1}v_{1} \), and 
the rest of the \( \lambda _{ij}^{q} \) are zero.
The values in parentheses correspond to the bottom quark/squarks. The \( k_{i}^{q} \)
and \( l_{i}^{q} \) are defined above, following Eq. (\ref{Eq11}). 
The functions \( {\mathcal{F}} \) and \( {\mathcal{G}} \) that appear
in Eq. (\ref{Eq15}), are the usual loop amplitudes which also appear
in the MSSM effective potential. These are given by \begin{eqnarray}
{\mathcal{F}}\left( x,y\right)  & = & \ln \left( \frac{xy}{Q^{4}}\right) +\frac{y+x}{y-x}\ln \left( \frac{y}{x}\right) -2,\nonumber \\
{\mathcal{G}}\left( x,y\right)  & = & \ln \left( \frac{xy}{Q^{4}}\right) -{\mathcal{F}}(x,y),\label{Eq18} 
\end{eqnarray}
 where one particularly notices that \( {\mathcal{F}} \) depends
explicitly on the renormalization scale. By combining Eqs. (\ref{Eq6}), (\ref{Eq10}),
and (\ref{Eq14}), we can diagonalize the total pseudoscalar mass-squared
matrix to obtain the mass of the pseudoscalar: 
\begin{eqnarray}
M_{P}^{2} & = & \left( M_{P}^{0}\right) ^{2}\left[ 1+\sum_{q=t,b}\frac{3h_{q}^{2}}{32\pi ^{2}}\frac{A_{q}}{A}{\mathcal{F}}\left(
m_{\tilde{q}_{1}}^{2},m_{\tilde{q}_{2}}^{2}\right) \right].\nonumber \\
\label{Eq19} 
\end{eqnarray}
We can check the validity of our expressions for \( M_{P}^{2} \)
and \( {\mathcal{M}}_{ij} \) by comparing them to the well-known
MSSM results. Our model reduces to MSSM if we fix \( \mu _{s}=h_{s}x/\sqrt{2}\equiv \mu  \)
and set \( h_{s}=g_{1}'=Q_{i}=0 \). In this limiting case, we identically
recover the usual MSSM results computed in \cite{ellis2}. We can also obtain a useful upper bound 
for the lightest Higgs mass at the one-loop level.  
\begin{eqnarray}
M_{h_{1}}^{2} & \leq  & \left( M_{h_{1}}^{0}\right) _{max}\nonumber\\
& + & \cos ^{2}\beta({\mathcal{M}}_{S}^{1})_{11}
+\sin 2\beta({\mathcal{M}}_{S}^{1})_{12}+\sin ^{2}\beta({\mathcal{M}}_{S}^{1})_{22}.\nonumber\\
\label{one loop mass upper bound} 
\end{eqnarray}
Although we did not
take into account the two-loop effects, we do not expect these corrections
to be very large. Indeed, as was shown in Ref. \cite{espinosa} for MSSM, the two-loop
effects are less than a few GeVs. In our future work, we will include the
two-loop effects in the model under consideration. But given the fact the not
even the one-loop effects have been worked out in this model, it is important
to have these corrections first before higher order effects are taken into
account.

\section{Numerical Examples}

We numerically diagonalize the total one-loop scalar mass-squared
matrix, given by Eq. (\ref{Eq15}), to obtain the mass of the lightest
Higgs scalar. In order to get concrete results, we must fix some of
the parameters in our model, which include \( g_{2} \), \( g_{1}' \),
\( h_{t} \), \( h_{s} \), \( m_{U} \), \( m_{D} \), \( m_{Q} \), \( A_{t} \), \( A_{b} \),
\( A \), \( v \), \( x \), \( \tan \beta  \), \( Q_{1} \) and
\( Q_{2} \) . The \( U(1)' \) gauge symmetry does not affect the
mass of the \( W \) boson. This fixes \( g_{2}=0.65 \) and \( v=246 \)
\( {\textrm{GeV}} \). According to the one-loop RGE analysis of \cite{cvetic},
\( h_{t}\sim 1.0-1.2 \) and \( 0.35 \leq h_{s} \leq 0.9-1.0 \). We leave the soft 
supersymmetry breaking masses \( m_{Q,U,D} \) and \( A \), \( A_{t,b} \),
\( h_{s} \), \( \tan \beta \), and \( Q \) as free parameters against which
the lightest Higgs mass will be plotted. Furthermore, because we are
considering the electroweak symmetry breaking driven by a large VEV
of the singlet field \( S \), much larger than \( v \), we fix \( x=1.5 \) TeV.

The values of \( g_{1}' \), \( Q_{1} \) and \( Q_{2} \) are not
directly constrained by the experimental data. However, in most GUT-motivated
models with an extra \( U(1)' \) factor, 
\( g_{1}'=\left( \sqrt{5\lambda _{g}/3}\right) g_{1}\tan \vartheta _{w} \),
where \( g_{1} \) is the hypercharge gauge coupling, and 
\( \lambda _{g} \) is of the order of one with the exact value
depending on how the GUT gauge group is broken down to the SM gauge
group \cite{robinett}. For our calculations, we take \( g_{1}' = 0.3 \).
To fix the charges \( Q_{i} \), we take the \( Z-Z' \) mixing angle
\( \alpha _{Z-Z'} \) to be smaller than a few times \( 10^{-3} \). Without excessive
fine-tuning of the theory, this requires \( h_{s}\sim g_{1}'Q_{S} \)
(see \cite{cvetic} for details). This together with \( 0.35 \leq h_{s} \leq 0.9 \)
imply \( Q_{s}\sim O(1) \). Furthermore, the gauge invariance of
the superpotential under the \( U(1)' \) gives \( Q_{1}+Q_{2}=-Q_{s} \).
Because \( Q_{1}Q_{2}>0 \) as is argued in \cite{cvetic}, we conclude
that \( \left| Q_{i}\right| \sim 1 \). A natural choice would be
to take \( Q_{1}=Q_{2}=-1 \). 

However, according to Eq. (\ref{Eq9}),
this specific choice restricts the value of \( \tan \beta  \) severely.
In fact, \( \alpha _{Z-Z'}\leq 3\times 10^{-3} \) implies \( 1\leq \tan \beta \leq 1.3 \).
It turns out that \( \tan \beta  \) is highly sensitive to the ratio
of the charges. Fortunately, the masses of the Higgs bosons are not
very sensitive to the charges at all. This is because in the mass-squared
matrix, each charge \( Q_{i} \) is accompanied by a factor of \( g_{1}' \).
At the one-loop level, it is the top quark Yukawa coupling that gives
the largest contribution to the Higgs masses. Therefore, we are free
to choose \( Q_{i} \) without effecting our results significantly.
We have also verified this numerically. Because \( \tan \beta  \)
is the only quantity that is really sensitive to the ratio of the
charges, we choose charges that allow \( \tan \beta  \) to vary over
a wide range of values. If we take \( Q_{1}=3Q_{2}=-3/2 \), then
we can have \( 1\lesssim \tan \beta \lesssim 60 \) without violating
the bound on \( \alpha _{Z-Z'} \). If we choose other values for
the charges that are still of the order of one, the masses of the Higgs bosons
would not change by a significant amount.

We find that the Higgs masses are sensitive to the value
of \( h_{s} \). For this reason, we calculate the Higgs masses for
\( h_{s}=0.4 \), \( 0.6 \), and \( 0.8 \). Finally, we fix the
scale at which we carry out the numerical calculations. We take the
scale, denoted by \( Q \) not to be confused with the charges \( Q_{i} \),
to be of the order of the electroweak scale. This is indeed necessary
for the consistency of our analysis. We have also verified that as
\( Q \) varies between \( 100-2000 \) GeV, the Higgs masses
change by less than \( 2-3 \) GeV almost independently of
the other parameters, see Fig. \ref{fig2}(c). It is clear that the change in the lightest Higgs mass
is because of different choices of \( h_{s} \), and not because of \( Q \). 
That the Higgs masses are relatively stable against
the choice of the scale, is a restatement of the stability of the
effective potential against the scale when one-loop effects are taken
into account, as we mentioned above. In the actual calculations we
fix \( Q=300 \) GeV. 

In summary, we are studying the lightest Higgs mass, which is phenomenologically
the most interesting one, as we vary \( m_{U,D,Q} \), \( A_{t,b} \), \( A \), \( \tan \beta  \),
and \( h_{s} \). We have summarized
our results in Figs. \( 1 \) and \( 2 \). We have plotted
the lightest Higgs mass for \( 100\leq A\leq 2000 \) GeV, \( -1600\leq A_{t,b} \leq 1600 \) GeV,
\( 200\leq m_{U,D,Q}\leq 1000 \) GeV, \( 1\leq \tan \beta \leq 60 \), and \( 0\leq h_{s} \leq 1.0 \).
We take \( m_{U,D,Q}\geq 200 \) GeV to keep squark masses positive. A negative mass for even
one of the squarks induces color breaking minima in the potential, which is clearly undesired. The above bounds
provide a fairly comprehensive region for our parameters. The physically allowed values 
must respect the LEPII constraint, \( M_{h_{1}} \geq 114 \) GeV \cite{lep}. 

Using Figs. \ref{fig1} and \ref{fig2}, we summarize our main results as follows:
First, \( h_{s} \geq 0.8 \) is excluded if \( \tan \beta \geq 2 \), see Fig. \ref{fig2}(e). For \( \tan \beta \leq 2 \),
\( h_{s} \) can be \( 0.8 \) or larger. However, according to Figs. \ref{fig1}(a-f), when \( h_{s} \geq 0.8 \), 
\( \tan \beta \geq 4 \) is excluded unless \( A \geq 1300-1350 \) GeV. We conclude that \( h_{s} \geq 0.8 \) and 
\( \tan \beta \geq 4 \) cannot be allowed simultaneously.
Second, the lightest Higgs mass, \( M_{h_{1}} \), is relatively stable
against the soft supersymmetry breaking masses \( m_{U,D,Q} \) and against \( A_{b} \), see Fig. \ref{fig1}(a,b,c,e).
There is some variation with respect to \( m_{Q} \), \( m_{D} \), and \( A_{b} \), but only when 
\( \tan \beta \) is very large. \( M_{h_{1}} \) varies quite a bit when \( A_{t} \) and \( A \) vary.
However, when \( \tan \beta \) is not near its extreme values, either too large or close to \( 1 \),
then \( M_{h_{1}} \) is relatively stable against \( A_{t} \) and \( A \) as well, see Fig. \ref{fig1}(d,f).
\( M_{h_{1}} \) stays pretty much fixed as \( Q \) varies between \( 200-2000 \) GeV. This is demonstrated in 
Fig. \ref{fig2}(c), for different values of \( h_{s} \). We have also plotted the lightest Higgs
mass as a function of \( \mu_{s} \), Figs. \ref{fig2}(a,b). Notice the extreme fine-tuning
of the charges \( Q_{1} \) and \( Q_{2} \) in order to have \( \tan \beta \) large with no
restriction on \( x \), and respecting the experimental bound on the \( Z-Z^\prime \) mixing angle.
With Fig. \ref{fig2}(b), we arrive at the same conclusion as above.
Namely, for large \( \tan \beta \), \( h_{s} \leq 0.8 \). Larger values of \( \tan \beta \), however,
are not generally favored by the above model.

We have verified numerically that for nominal values of \( \tan \beta \) and \( h_{s} \), 
say, \( \tan \beta \sim 4-15 \) and \( h_{s} \sim 0.4-0.6 \), the lightest Higgs mass is about 
\( 130 \) GeV, which is very close to the MSSM prediction. This can be seen from our graphs by 
examining the dashed black and red curves. Based on the MSSM analysis \cite{espinosa}, we do not 
expect the two-loop corrections to add more than a few GeVs to this value. That there is an agreement
between the above model and MSSM is no surprise, because the above model is just an extension
of MSSM. This is in fact a check on the validity of our results. 

The above conclusions might be slightly effected when CP-violating effects are taken into account.
However, the inclusion of CP-violating phases coming from \( h_{s}A_{s} \)
and \( h_{t}A_{t} \) mixes the scalar and pseudoscalar mass-squared
matrices. The resulting mass-squared matrix is a \( 4\times 4 \)
matrix giving four Higgs bosons with no definite CP properties. This
requires a different analysis than what we have done in this paper.
We consider CP violation in a separate work. 

Finally, for phenomenological reasons, it is desirable to know the
Higgs mixing angles. In Fig. \( 2 \)(f), we have plotted the three Higgs mixing
angles against \( \tan \beta  \). The derivation of the Higgs mixing
angles is shown in appendix A, which is already well known. From the
graph, one can see that for smaller values of \( \tan \beta  \),
the lightest Higgs state is dominated by a mixture of \( H_{1}^{0} \)
and \( H_{2}^{0} \), while its mixing with the singlet stays less
than about \( 10 \) percent. However, when \( \tan \beta  \) is larger than
about \( 10 \), the lightest state is dominated by \( H_{2}^{0} \),
and the total mixing with \( H_{1}^{0} \) and \( S \) stays less
than \( 10 \) percent. This result is not surprising because larger
\( \tan \beta  \) implies a larger VEV for \( H_{2}^{0} \).

\section{Conclusions}

In this work, we studied the one-loop effects on the lightest Higgs
mass in a minimal supersymmetric model augmented by an Abelian \( U(1)' \)
gauge symmetry. We calculated the top and stop/sbottom one-loop effects in
the framework of the effective potential approach. 
The most important issue concerning the minimal
models extended by a \( U(1)' \) factor, is the mixing of the Standard
Model \( Z \) boson with the \( Z' \) boson associated with the
extra gauge symmetry. In order for such models to be phenomenologically
viable, the \( Z \)--\( Z^{\prime } \) mixing angle has to be very
small, less than a few times \( 10^{-3} \). We used the smallness of the
\( Z \)--\( Z^{\prime } \) mixing angle, together with the lower
bound set on the lightest Higgs mass, \( 114 \) GeV, by LEPII data \cite{lep}, to constrain
our parameter space. We showed numerically that the one-loop effects
due to the top and stop/sbottom quarks are non-negligible.

The radiative corrections to the Higgs boson masses and mixing angles are
crucial for interpreting and predicting the Higgs production and decay
rates in upcoming colliders. In linear colliders, for example NLC,
the main production mechanisms are the Bjorken process and pair--production
process, each of which requires a precise knowledge of Higgs boson
masses and their couplings to the gauge bosons. (For the analysis
of these processes at the tree level, see \( e.g \) \cite{durmush}).
The model at hand predicts a larger upper bound on the Higgs boson
masses than MSSM. Therefore, even if the MSSM bounds are violated
in the near-future colliders, the model at hand, which generates the
\( \mu  \) parameter dynamically, will accommodate larger Higgs masses.

\section{Acknowledgement}

We would like to thank D.A. Demir and A. Vainshtein for useful discussions. This
work was supported by the University of Minnesota under the Doctoral
Dissertation Fellowship grant.

\appendix
\section{Derivation of the Higgs Masses and Mixing Angles}

The Higgs mass-squared matrix is given by a symmetric \( 3\times 3 \)
matrix, \( \left( {\mathcal{M}}_{ij}\right)  \). The eigenvalues
are given by the solutions of the following characteristic equation:
\( x^{3}+rx^{2}+sx+t=0 \), where

\begin{eqnarray}
r & = & -tr\left( {\mathcal{M}}_{ij}\right) ,\nonumber \\
s & = & -\frac{1}{2}\left[ tr^{2}\left( {\mathcal{M}}_{ij}\right) -tr\left( {\mathcal{M}}_{ij}^{2}\right) \right] ,\nonumber \\
t & = & -det\left( {\mathcal{M}}_{ij}\right) .
\end{eqnarray}
 With the help of the auxiliary parameters \( p=s-r^{2}/3 \), \( q=2r^{3}/27-rs/3+t \),
\( d=p^{3}/27+q^{2}/4 \), and \( \vartheta =\arccos \left( -q/\sqrt{-4p^{3}/27}\right)  \),
we can express the Higgs masses as follows:

\begin{eqnarray}
M_{h_{1}}^{2} & = & -\frac{1}{3}r+2\sqrt{-\frac{p}{3}}\cos \left( \frac{\vartheta }{3}+\frac{2\pi }{3}\right) ,\nonumber \\
M_{h_{2}}^{2} & = & -\frac{1}{3}r+2\sqrt{-\frac{p}{3}}\cos \left( \frac{\vartheta }{3}-\frac{2\pi }{3}\right) ,\nonumber \\
M_{h_{3}}^{2} & = & -\frac{1}{3}r+2\sqrt{-\frac{p}{3}}\cos \left( \frac{\vartheta }{3}\right) ,
\end{eqnarray}
 where we require \( d,r,t<0 \) and \( s>0 \), to ensure that the
masses are physical, i.e., positive. Because the Higgs mass-squared
matrix is real and symmetric, it can be diagonalized by means of an
orthogonal transformation \( {\mathcal{O}} \), where

\begin{eqnarray}
{\mathcal{O}}^{T}{\mathcal{M}}_{ij}{\mathcal{O}} & = & diag\left( M_{h_{3}}^{2},M_{h_{2}}^{2},M_{h_{1}}^{2}\right) .
\end{eqnarray}
 In the basis \( \left( \phi _{1},\phi _{2},\phi _{3}\right) =\left( H_{1}^{0},H_{2}^{0},S\right) \),
the mass eigensta mtes are defined by \( h_{4-i}=\sum _{j=1}^{3}{\mathcal{O}}_{ij}\phi _{j} \),
for \( i=1,2,3 \). More specifically, we have

\begin{eqnarray}
h_{1} & = & {\mathcal{O}}_{31}\phi _{1}+{\mathcal{O}}_{32}\phi _{2}+{\mathcal{O}}_{33}\phi _{3},\nonumber \\
h_{2} & = & {\mathcal{O}}_{21}\phi _{1}+{\mathcal{O}}_{22}\phi _{2}+{\mathcal{O}}_{23}\phi _{3},\nonumber \\
h_{3} & = & {\mathcal{O}}_{11}\phi _{1}+{\mathcal{O}}_{12}\phi _{2}+{\mathcal{O}}_{13}\phi _{3}.
\end{eqnarray}
 We are defining our fields such that \( M_{h_{1}}^{2}\leq M_{h_{2}}^{2}\leq M_{h_{3}}^{2} \).
Due to orthogonality, we have \( {\mathcal{O}}_{1i}^{2}+{\mathcal{O}}_{2i}^{2}+{\mathcal{O}}_{3i}^{2}=1 \),
for \( i=1,2,3 \). Since we are mainly interested in the mass of
the lightest Higgs boson, we only plot the values of \( {\mathcal{O}}_{31} \),
\( {\mathcal{O}}_{32} \), and \emph{\( {\mathcal{O}}_{33} \)}, as
these are the only mixing angles that determine the composition of
the lightest Higgs boson.

\begin{widetext}

\begin{figure}
\resizebox*{0.45\textwidth}{0.3\textheight}{\rotatebox{270}{\includegraphics{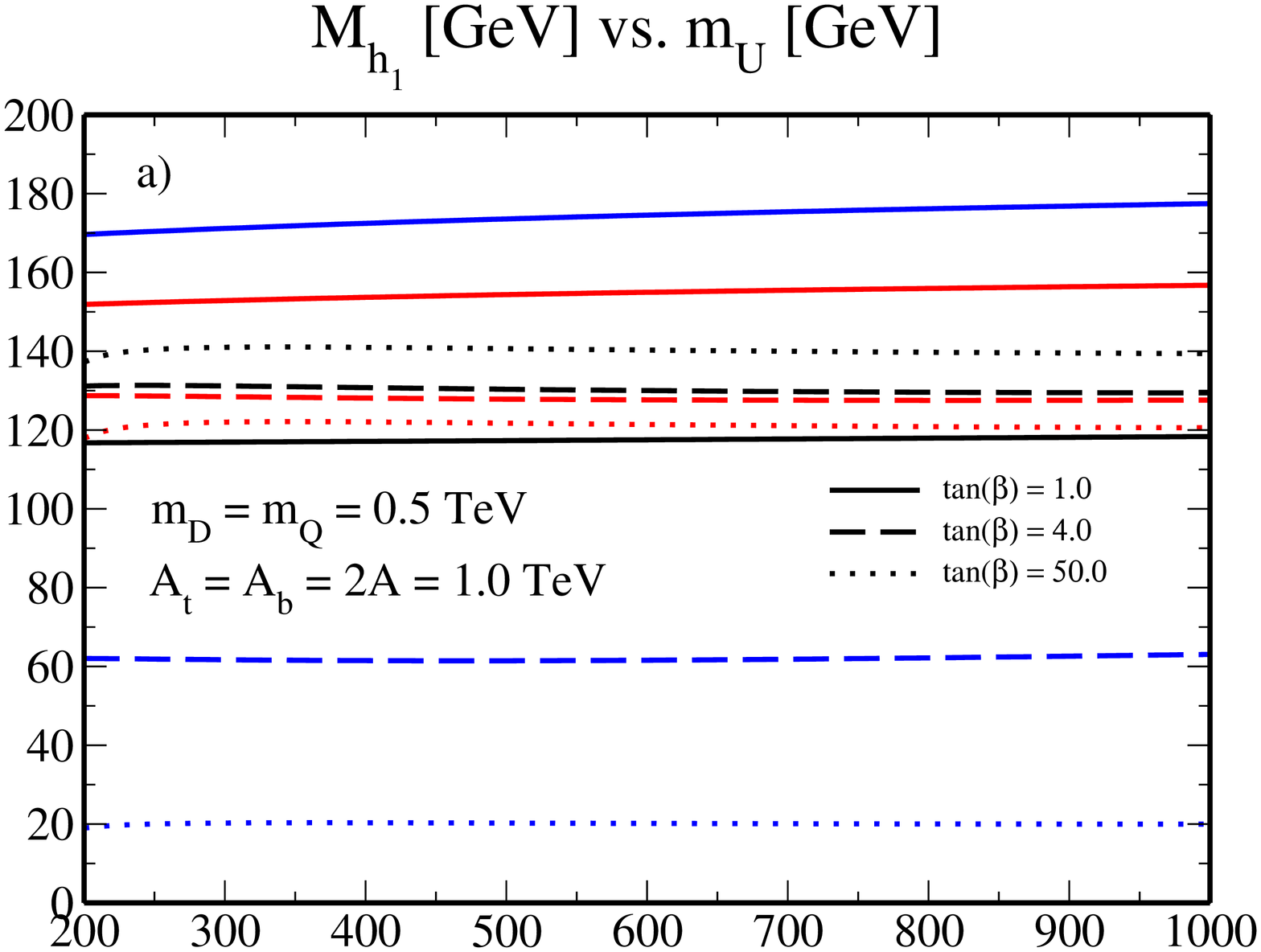}}}
\resizebox*{0.45\textwidth}{0.3\textheight}{\rotatebox{270}{\includegraphics{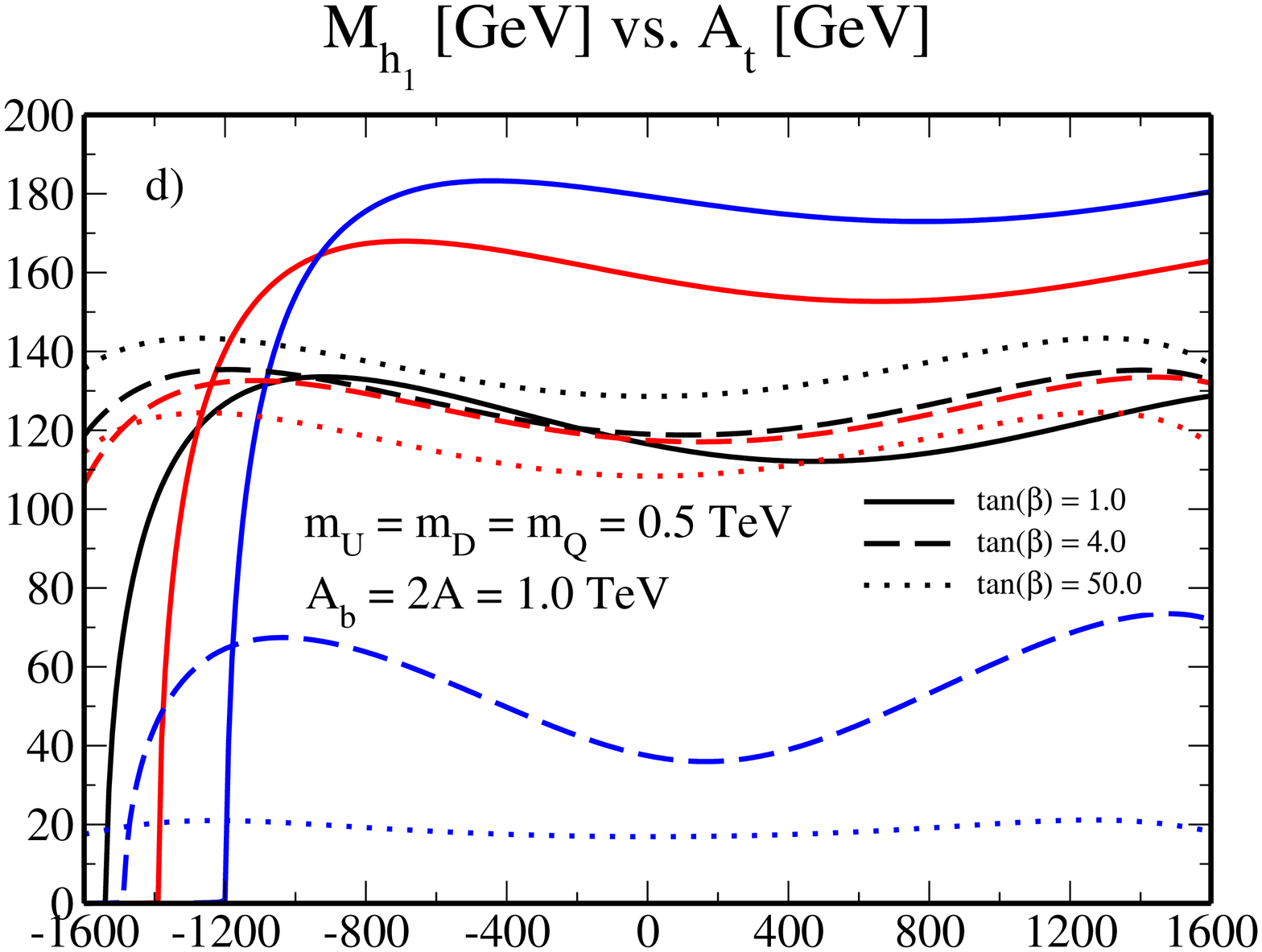}}}
\resizebox*{0.45\textwidth}{0.3\textheight}{\rotatebox{270}{\includegraphics{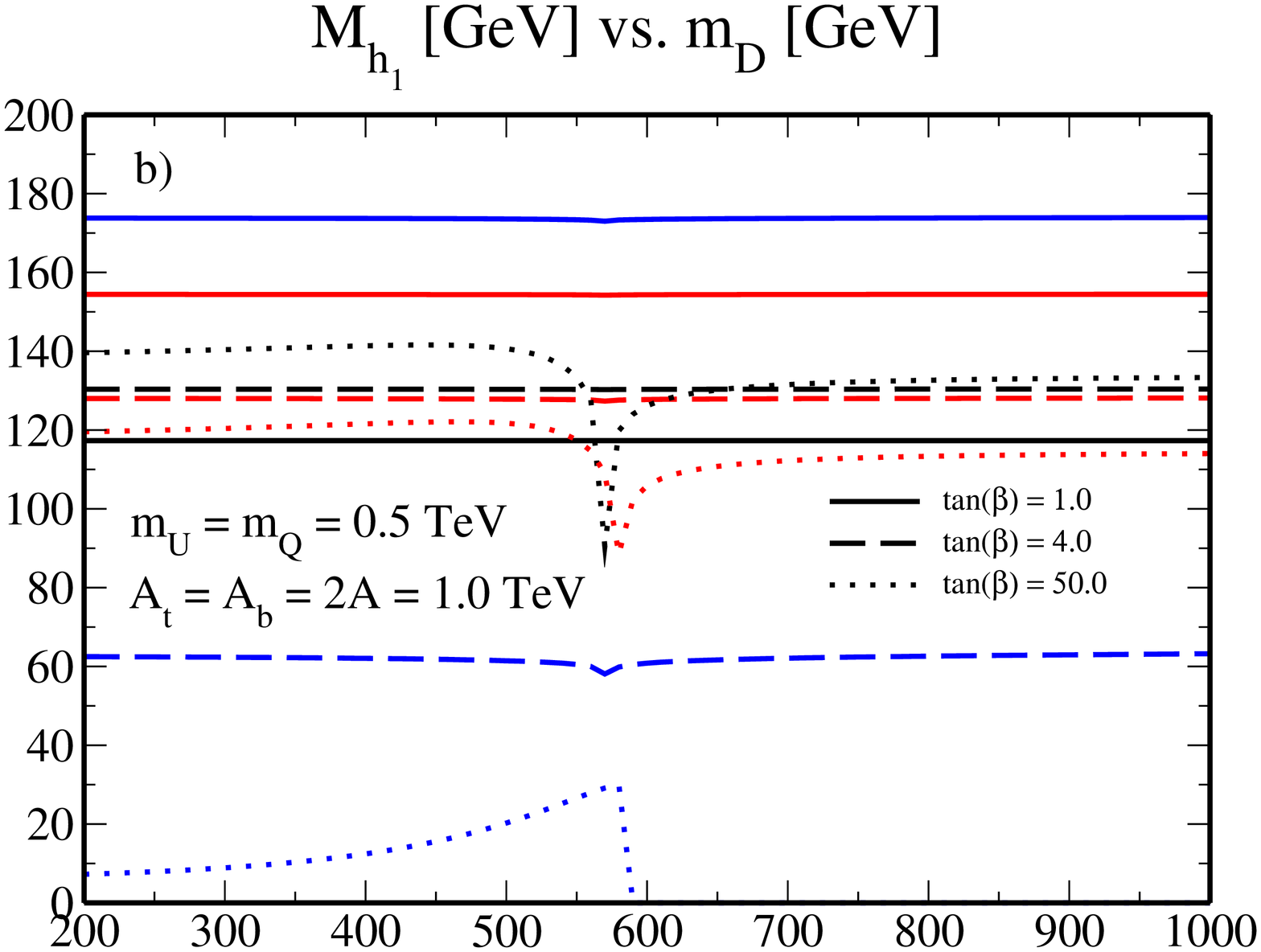}}} 
\resizebox*{0.45\textwidth}{0.3\textheight}{\rotatebox{270}{\includegraphics{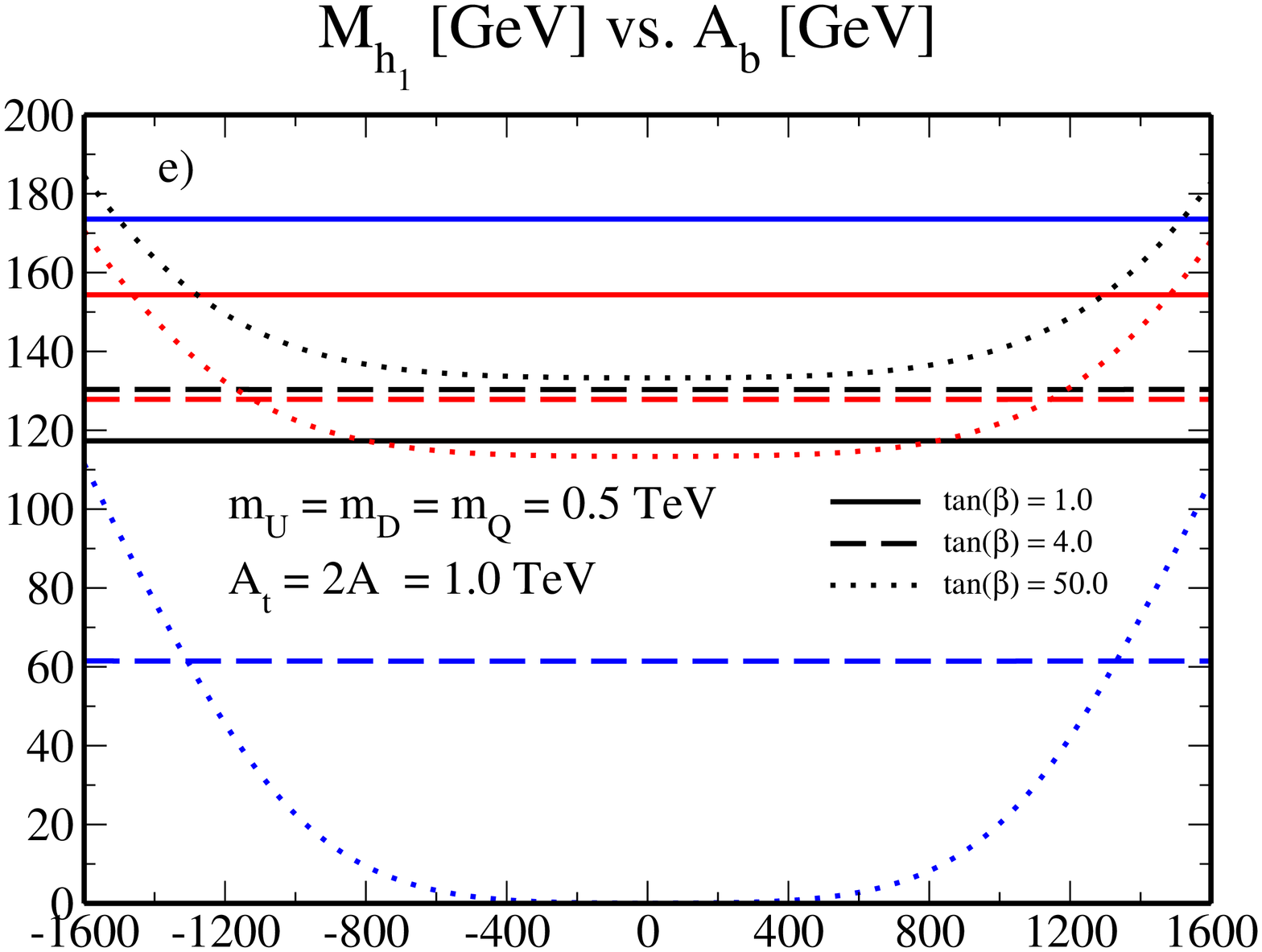}}}
\resizebox*{0.45\textwidth}{0.3\textheight}{\rotatebox{270}{\includegraphics{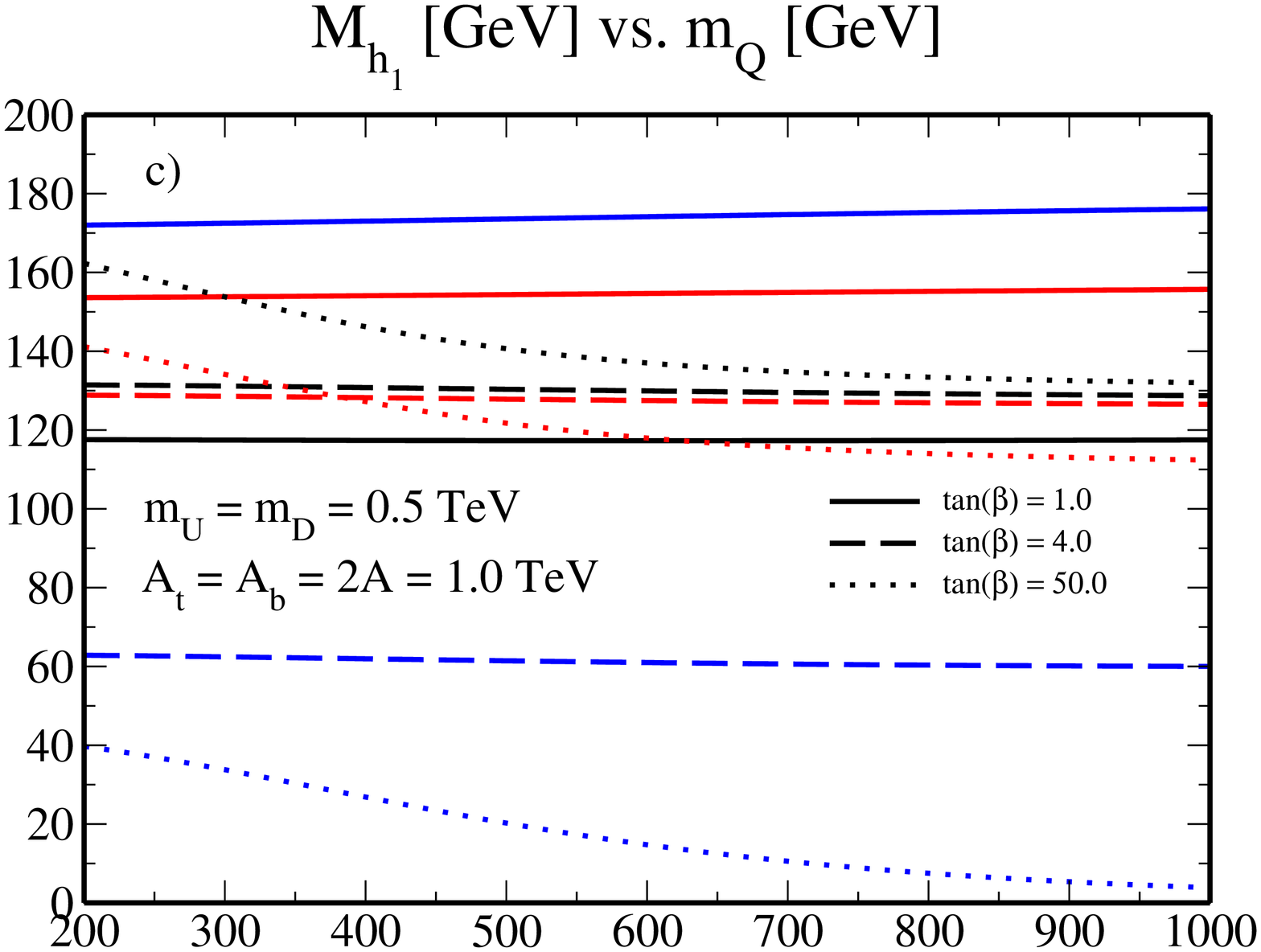}}} 
\resizebox*{0.45\textwidth}{0.3\textheight}{\rotatebox{270}{\includegraphics{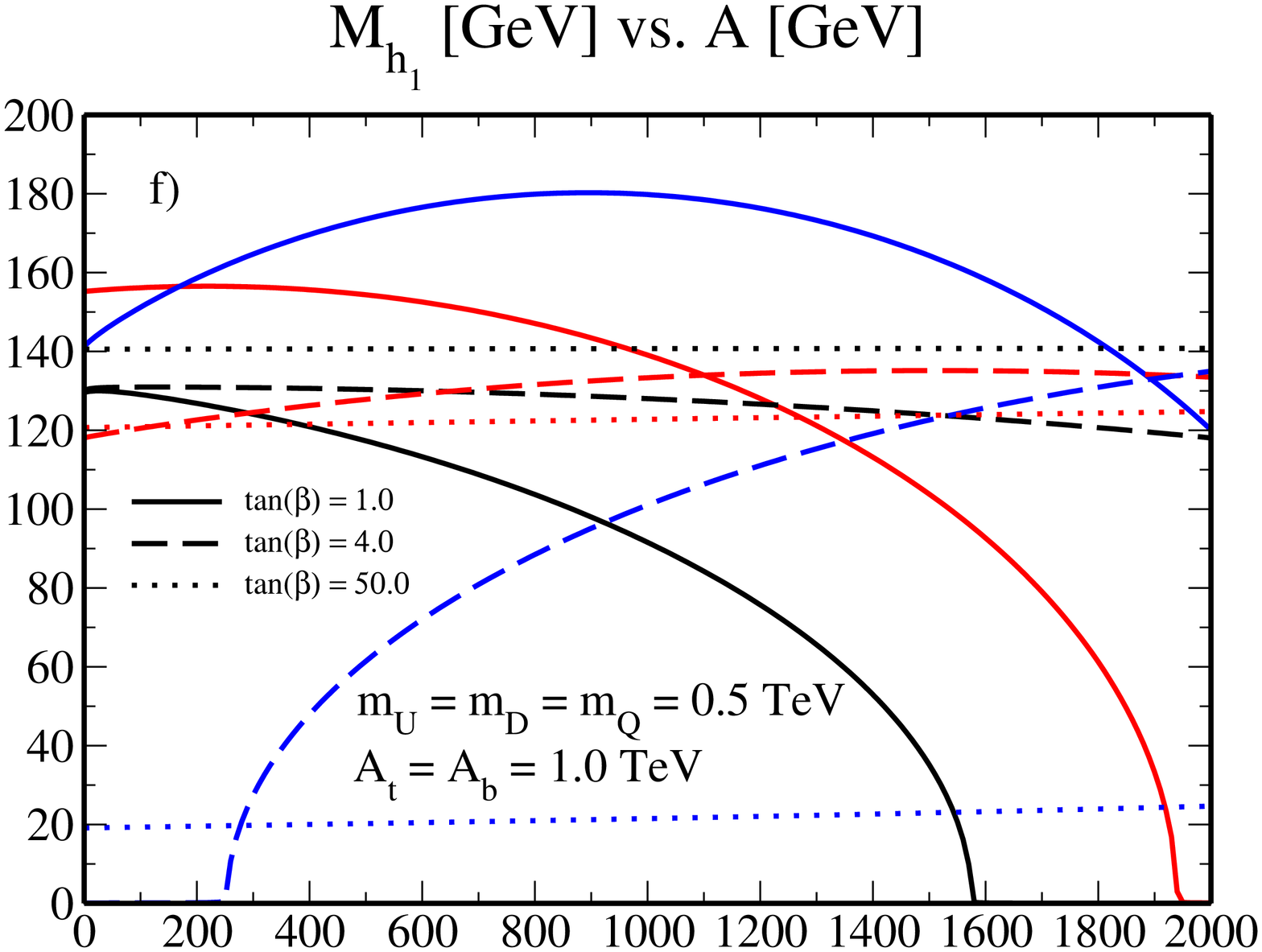}}}
\caption{Black, red, and blue curves correspond to \( h_{s} = 0.4 \), \( 0.6 \), and \( 0.8 \), respectively.}\label{fig1} 
\end{figure}

\begin{figure}
\resizebox*{0.45\textwidth}{0.3\textheight}{\rotatebox{270}{\includegraphics{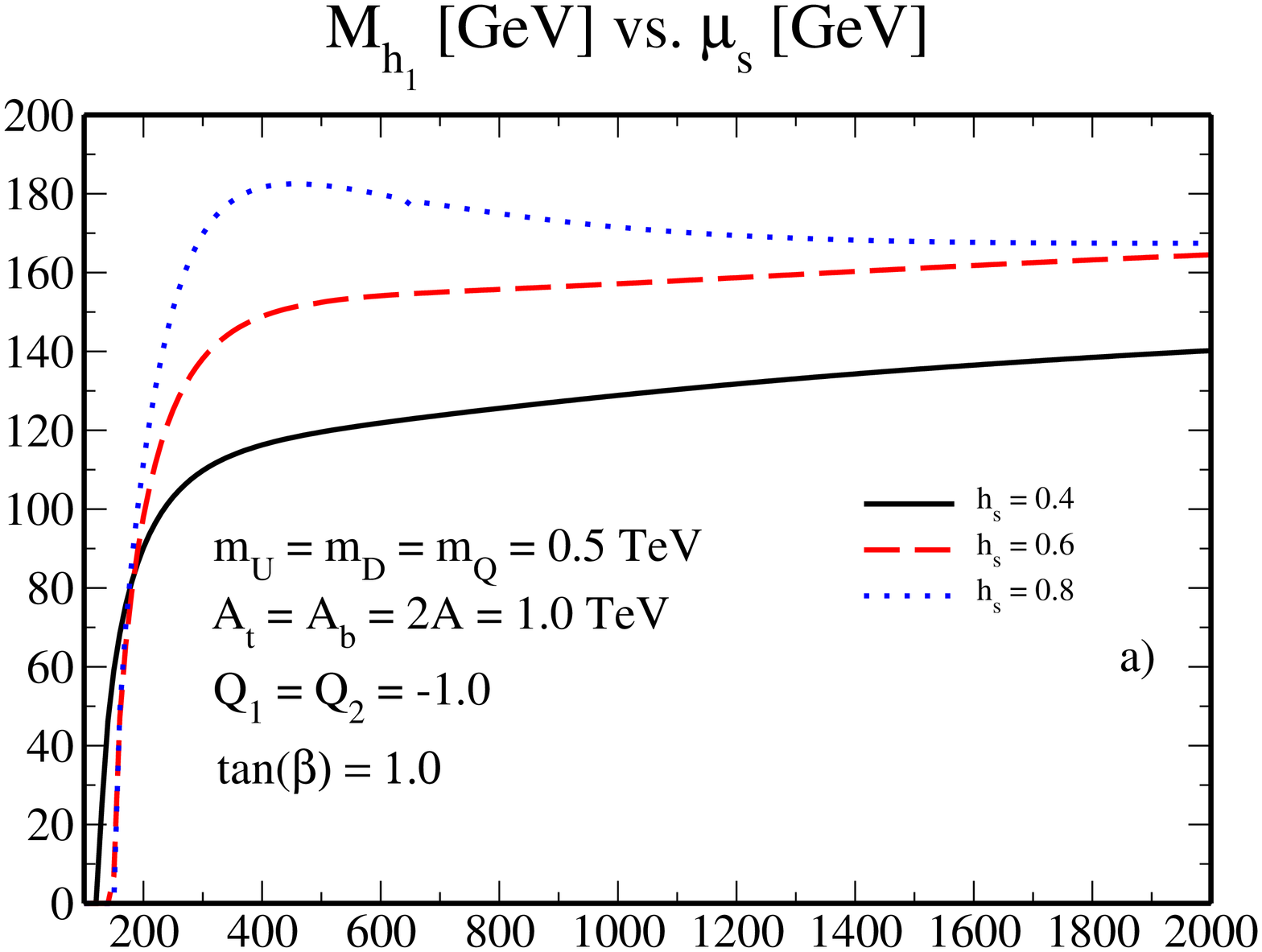}}} 
\resizebox*{0.45\textwidth}{0.3\textheight}{\rotatebox{270}{\includegraphics{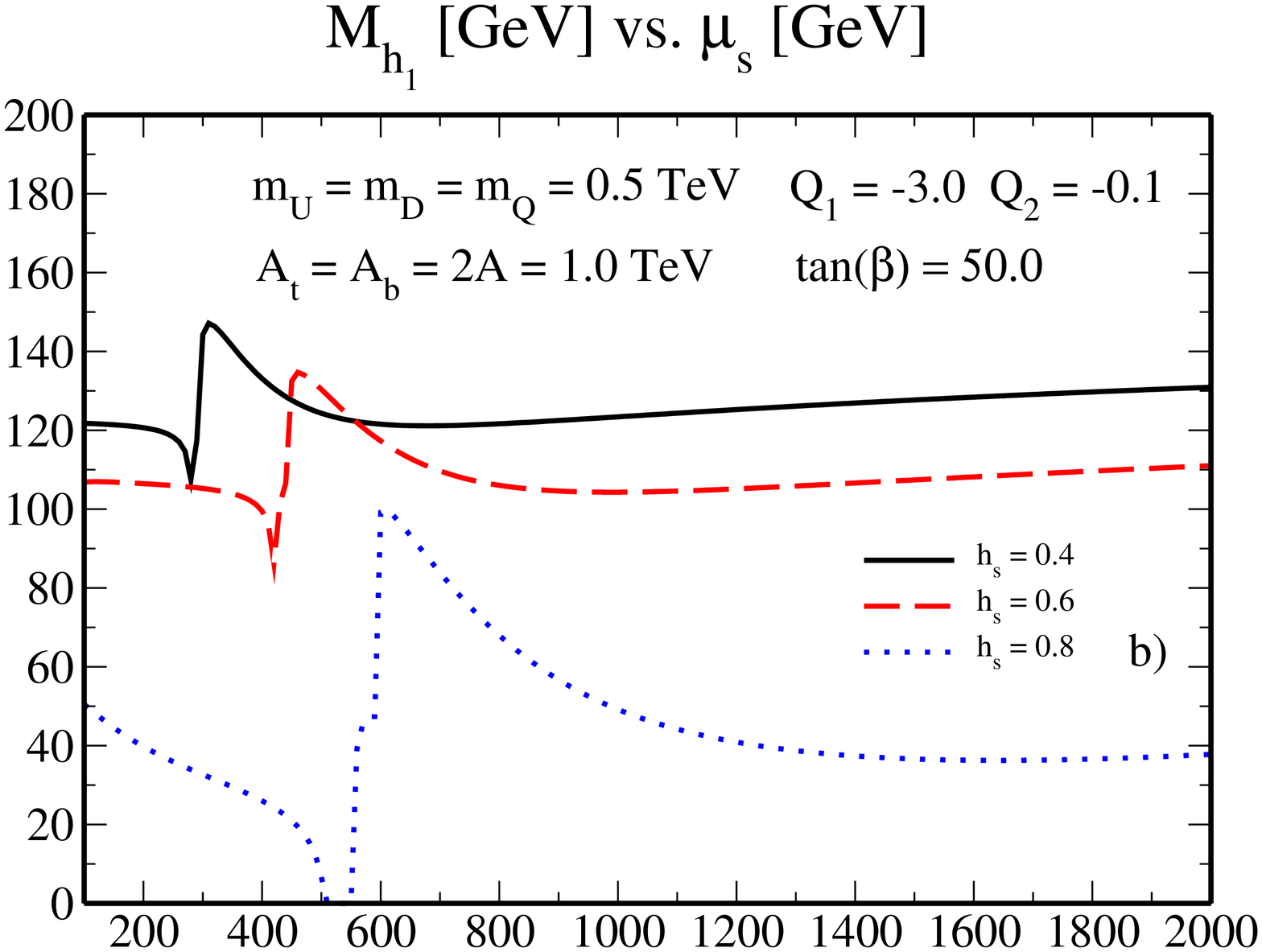}}} 
\resizebox*{0.45\textwidth}{0.3\textheight}{\rotatebox{270}{\includegraphics{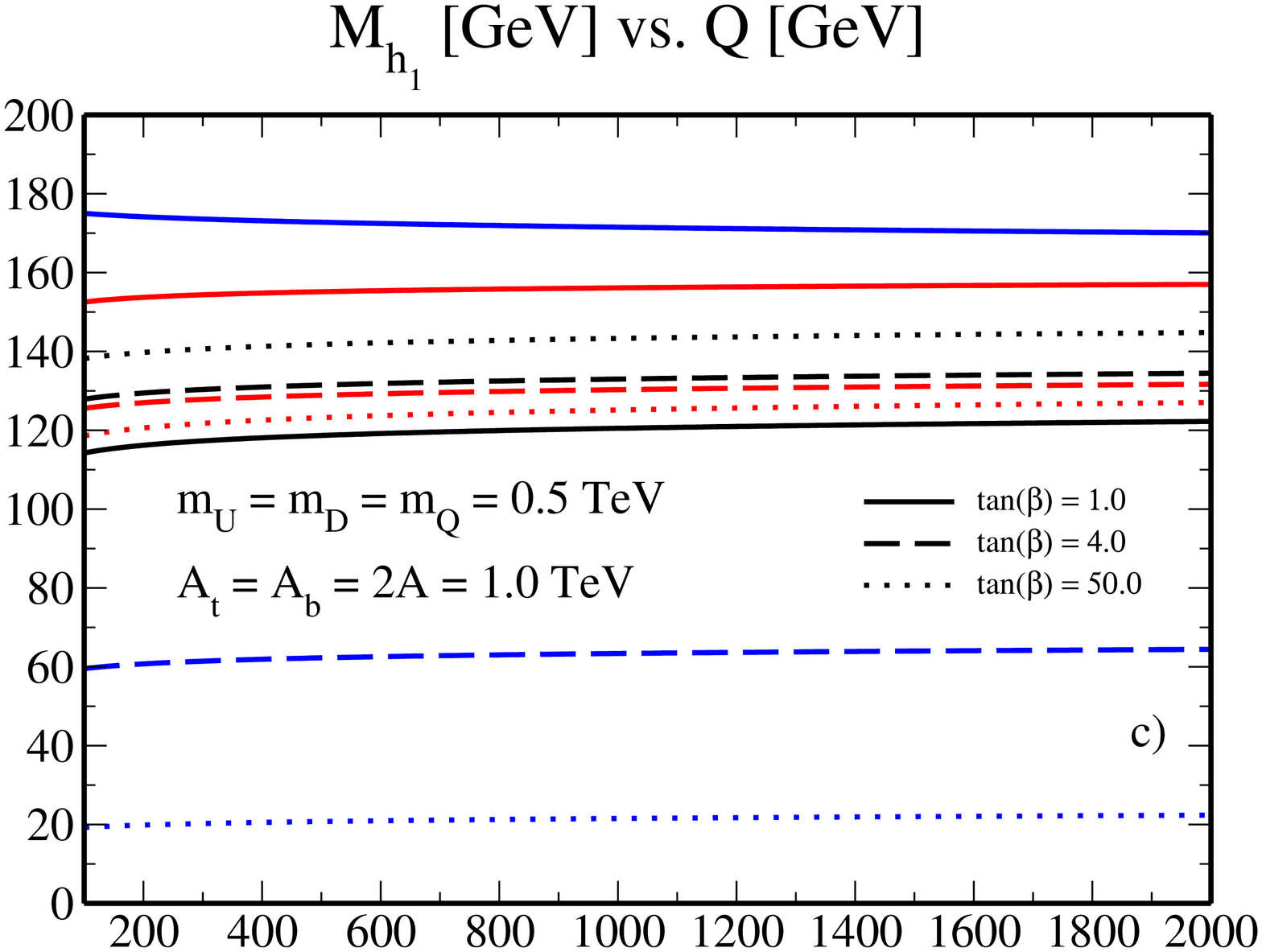}}} 
\resizebox*{0.45\textwidth}{0.3\textheight}{\rotatebox{270}{\includegraphics{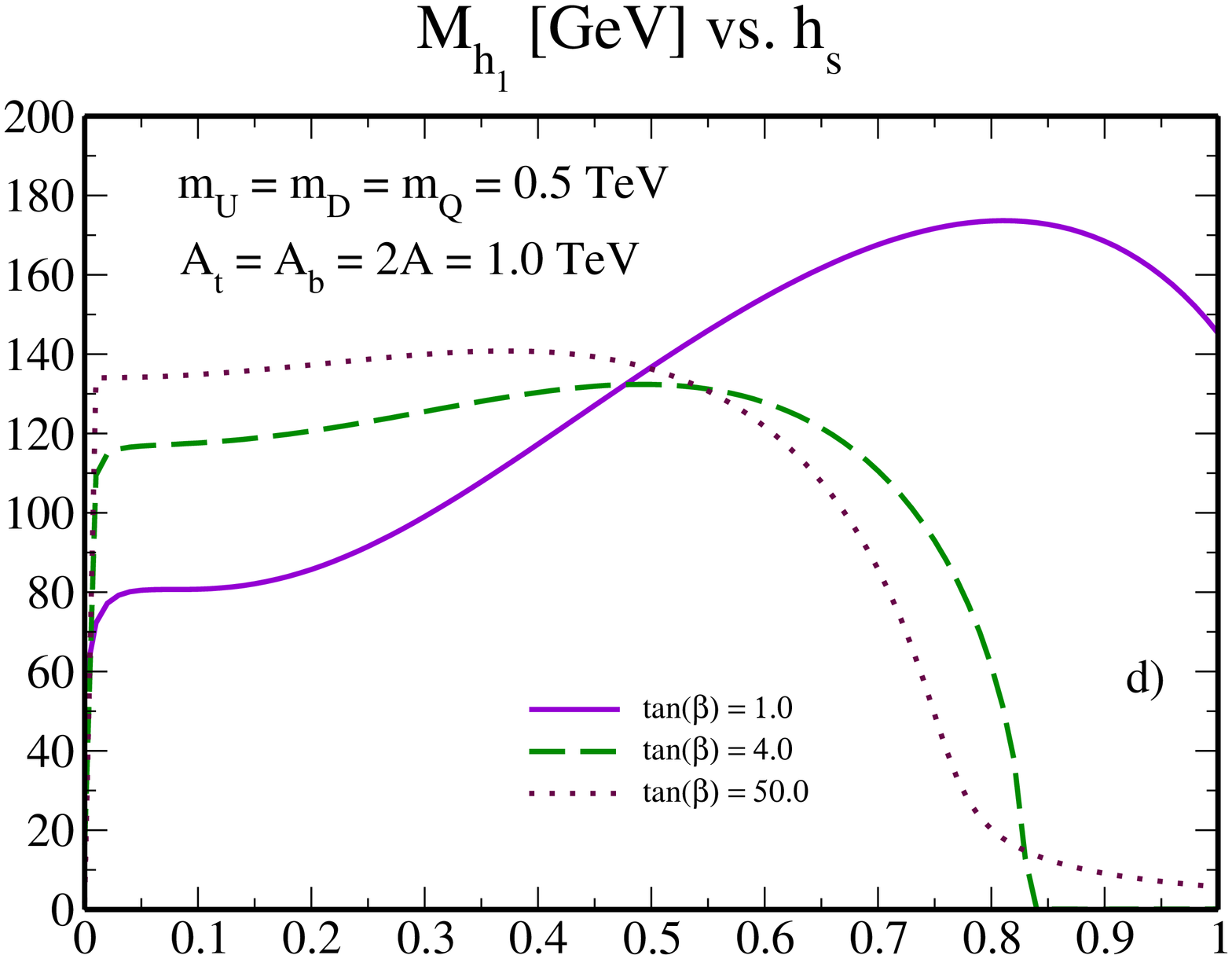}}} 
\resizebox*{0.45\textwidth}{0.3\textheight}{\rotatebox{270}{\includegraphics{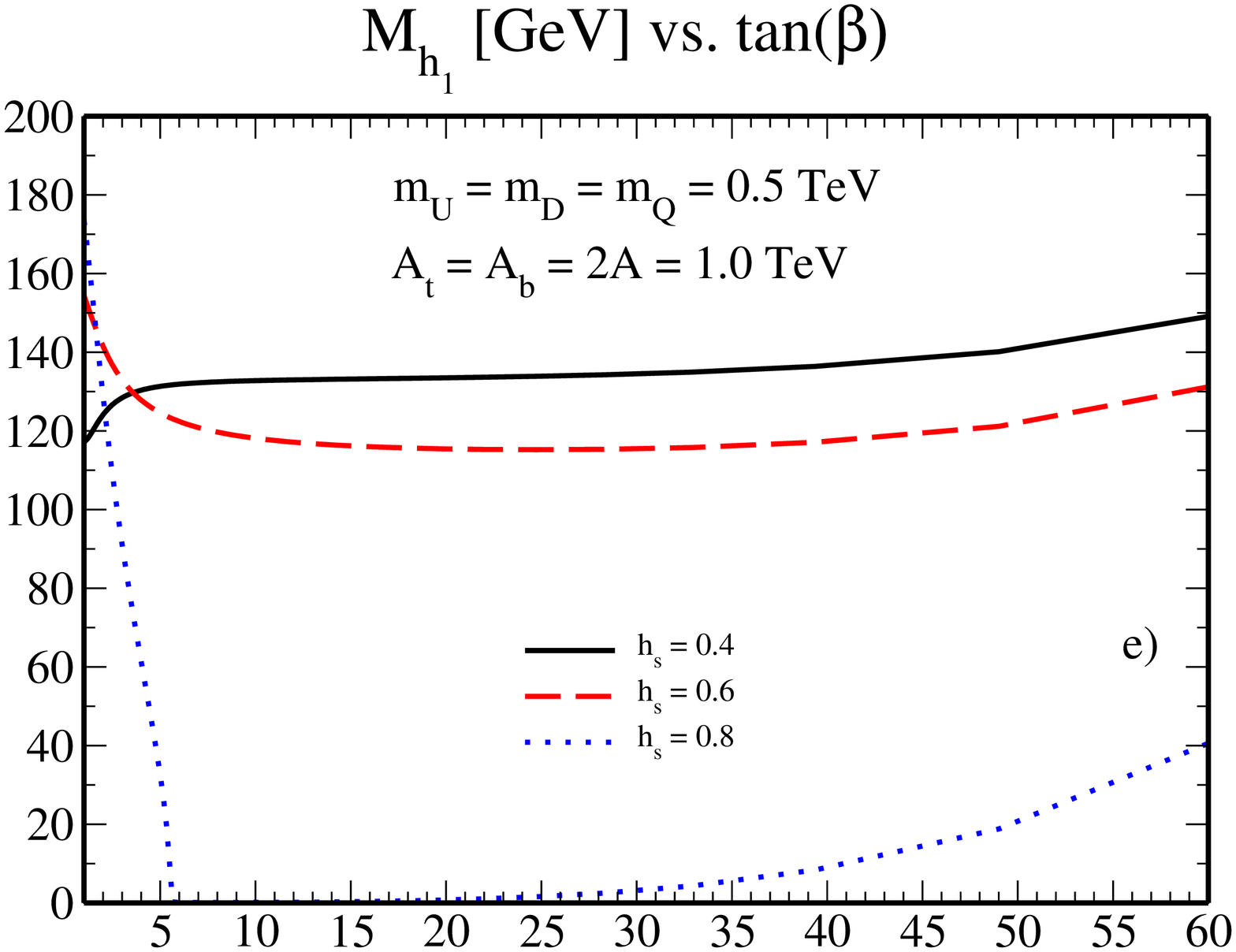}}} 
\resizebox*{0.45\textwidth}{0.3\textheight}{\rotatebox{270}{\includegraphics{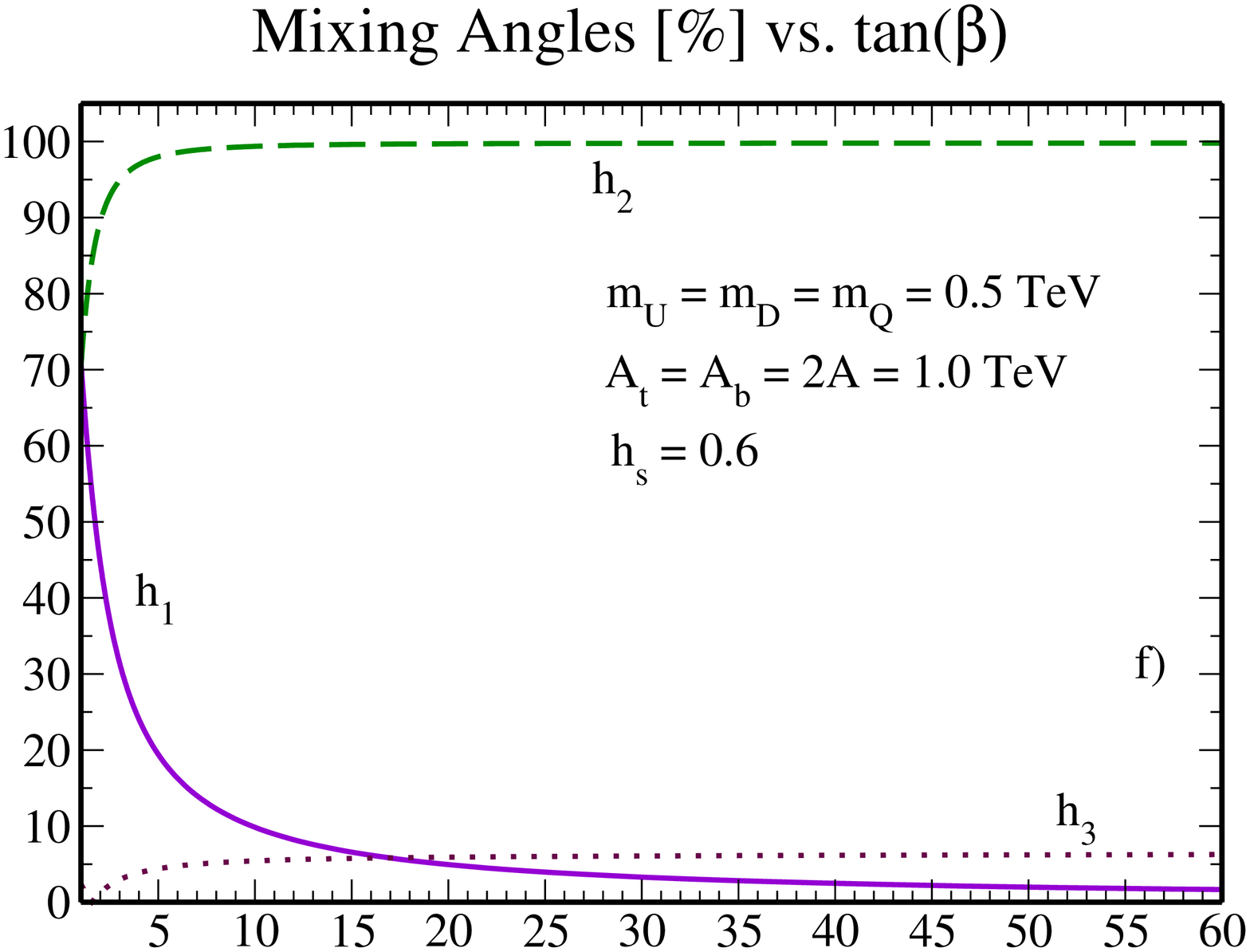}}}
\caption{Black, red, and blue curves correspond to \( h_{s} = 0.4 \), \( 0.6 \), and \( 0.8 \), respectively.}\label{fig2} 
\end{figure}
\end{widetext}


\begin{thebibliography}{10}
\bibitem{footnote}{In the usual NMSSM case, the presence of the \( \mu  \) 
term is needed in order to avoid the appearance of an axion after symmetry 
breaking. In our case, this axion becomes the longitudinal component of the
\( Z' \) associated with \( U(1)' \).}
\end{thebibliography}
\end{document}